\documentclass[aps,12pt,superscriptaddress,amsfonts,amssymb,amsmath]{revtex4}
\pdfoutput=1

\usepackage{graphicx}
\usepackage{epsfig}
\usepackage{makeidx}
\usepackage{epstopdf}
\usepackage{natbib}
\usepackage{xcolor}
\usepackage{subcaption}
\usepackage{amsmath}
\usepackage{appendix}

\begin{document}

\title{A new theory of tensor-scalar gravity coupled to Aharonov-Bohm electrodynamics
}
\author{F.\ Minotti \footnote{Email address: minotti@df.uba.ar}}
\affiliation{Universidad de Buenos Aires, Facultad de Ciencias Exactas y Naturales, Departamento de F\'{\i}sica, Buenos Aires, Argentina}
\affiliation{CONICET-Universidad de Buenos Aires, Instituto de F\'{\i}sica Interdisciplinaria y Aplicada (INFINA), Buenos Aires, Argentina}

\author{G.\ Modanese \footnote{Email address: giovanni.modanese@unibz.it}}
\affiliation{Free University of Bozen-Bolzano \\ Faculty of Engineering \\ I-39100 Bolzano, Italy}

\linespread{0.9}

\begin{abstract}
 
Tensor-scalar theories of gravitation are commonly employed as extensions of General Relativity that allow to describe a much wider phenomenology. They are also naturally generated as low energy limits of higher-dimensional or unified theories, and the gravitational scalar components can represent quantum corrections to the Einstein theory. The coupling of the scalars to an e.m. field does not introduce any relevant new physics if the e.m. action has the usual Maxwell form, implying a vanishing trace of the e.m. energy-momentum tensor. In the case of the extended Aharonov-Bohm electrodynamics some interesting new situations are possible, which in this work are analyzed in the gravitational weak-field approximation and for a basic version of tensor-scalar gravity involving only a Brans-Dicke field plus another scalar. Since Aharonov-Bohm theory differs from Maxwell theory only in the presence of anomalous sources with local violation of charge conservation, which is thought to be possible only at a quantum level, the resulting formal framework can be useful for modeling interactions between gravitational and physical systems with macroscopic quantization. The theory contains some unknown parameters, the most important being the vacuum expectation value (VEV) $\psi_0$ of the second gravitational scalar and the level $\gamma$ of violation of local charge conservation in the e.m. sector. An attempt is made to relate these parameters to some experimental constraints. However, presently there is much space left for uncertainty.
 
\end{abstract}

\maketitle

\section{Introduction}

In spite of the wide success of General Relativity (GR) as a comprehensive model of space-time and gravity, and of its beautiful logical and mathematical framework, over the last decades the emergence of some weak spots of the theory has motivated a search for possible extensions. It is now clear that GR is not the ultimate theory of gravitational interactions. Its problems arise mainly from cosmology and from quantum field theory, due to persistent difficulties in the quantization program. See for reviews \cite{capozziello2008extended,nojiri2011unified,nojiri2017modified}.

One of the oldest extensions of GR, the Brans-Dicke (BD) theory, introduces besides the metric tensor a scalar field non-minimally coupled to geometry. In the original intentions of its authors, the task of this field was to describe a variable $G$ incorporating Mach's principle. Due to the additional scalar field, the theory turns out to be compatible with a phenomenology wider than that of GR.

Unification schemes much more ambitious than the BD theory, like Kaluza-Klein, superstrings, supergravity and grand-unified theories \cite{birrell1984quantum,vilkovisky1992effective,gasperini1992d,buchbinder2017effective} reproduce GR as a low-energy limit, but it is a version of GR extended with scalar fields, and possibly with non-minimal couplings or terms in the action of higher order in $R$, compared to the Einstein action. Terms of higher order in $R$ are also generated by quantum corrections to the Einstein action, and it can be proven that via a conformal transformation each second-order derivative term corresponds to a scalar field \cite{gottlober1990sixth,ruzmaikina1970quadratic,amendola1993generalized,mayer1993sitter,schmidt1990variational}.

Therefore, one of the simplest possible models of extended gravity is given by an action which includes a non-minimally coupled scalar field of the BD kind, which we shall denote with $\phi$, and a second scalar $\psi$ which instead has a minimal coupling of the form $J\psi$, with $J$ defined in Eq. \eqref{ABsource} below. 

Note that in other approaches of this type, the field $\phi$ can be reabsorbed into a conformal transformation of the metric, see \cite{damour1992tensor}, with the advantage of making the $R$-term in the action equal to the Einstein term, and the disadvantage of introducing two metrics, of which only one is the physical metric. The latter convention is sometimes called the '' Einstein frame'', as opposed to the '' Fierz frame'' that we use here.

We are going to use this extended action as a reasonably simple but general starting point for coupling gravity to an extension of the Maxwell theory called Aharonov-Bohm electrodynamics, and for exploring the corresponding new phenomenology. Note that the scalar fields $\phi$, $\psi$ have vacuum expectation values (VEV) $\phi_0$, $\psi_0$. In order to preserve the usual GR limit, $\phi_0$ cannot be much different from 1, while the value of $\psi_0$ is formally defined by a self-interaction potential $U(\psi)$, which we do not specify here, and will in general be constrained by the cosmological phenomenology of the model, which is not considered in this work. We shall simply suppose that there exist non-vanishing values $\phi_0$, $\psi_0$, and we also treat as parameters of the theory, in the weak-field limit, the derivatives of the couplings $\beta_{mat}(\phi,\psi)$ and $\beta_{AB}(\phi,\psi)$ in the action (see below) with respect to $\phi$ and $\psi$, evaluated at $\phi_0$, $\psi_0$.

A tensor-scalar action similar to ours has been previoulsy studied, among others, by Mbelek, Lachi\`eze-Rey, Minotti and Raptis \cite{mbelek2002possible,mbelek2003five,mbelek2004modelling,minotti2013scalar,raptis2013effects,minotti2017revaluation}, in which the e.m. part of the action was that of Maxwell electrodynamics.
Those authors were mainly interested in the possible effects of the coupling between gravitational and e.m. fields which is present in this action and differs from the standard coupling in GR. They wondered whether this coupling could explain some interesting, though controversial claims like (1) a possible correlation between the measured values of the $G$ constant and the local strength of the Earth's magnetic field; (2) possible anomalous gravitational forces on resonant e.m. cavities.

That approach, however, is made difficult by one of the fundamental features of Maxwell theory, namely the fact that the trace of the e.m. energy-momentum tensor is vanishing: $\textrm{Tr}(T^{em}_{\mu\nu})=0$. Since the gravitational scalars are usually coupled to this trace, Mbelek and Lachi\`eze-Rey have instead suggested a coupling of the scalar $\psi$ to the e.m. Lagrangian density $F_{\mu\nu}F^{\mu\nu} \sim (\mathbf{E}^2-\mathbf{B}^2)$. The consequences of such a coupling are, unfortunately, very difficult to reconcile with some well-known phenomenological constraints \cite{minotti2013scalar}.

The perspective changes considerably in Aharonov-Bohm extended electrodynamics, which has only a reduced gauge invariance (\cite{EPJC2023} and refs.). In this extended theory, not only is the trace of $\textrm{Tr}(T^{em}_{\mu\nu})$ not generally zero, but there is another scalar quantity to which $\phi$ and $\psi$ can be naturally coupled, namely the e.m. scalar $S=\partial_\mu A^\mu$. This coupling is far less constrained by phenomenology, because one should consider that the source of $S$ in Aharonov-Bohm electrodynamics is the ''extra-current'' $I$, according to the equations $I=\partial_\mu j^\mu$, $\Box S=\mu_0 I$. Therefore, $S$ is thought to be non-zero only in very special situations, essentially due to quantum anomalies which spoil local charge conservation, or to macroscopic quantum fluctuations which introduce an indetermination in the local conservation relation $\partial_\mu j^\mu=0$ as a consequence of the phase-number uncertainty relation $\Delta N \Delta \Phi \ge \hbar$.

All this opens a new exciting possibility, completely absent in GR: a special coupling of the gravitational field with matter in a state which is able to generate an $S$-field. The special coupling is independent from the energy-momentum content $T^{mat}_{\mu\nu}$ of this matter, while $T^{mat}_{\mu\nu}$, on the contrary, is coupled to the metric independently from the state. About this general issue, namely the coupling of gravity to macroscopic quantum systems, only a relatively small number of studies exist, see e.g. the review \cite{gallerati2022interaction}.

One task of the extended tensor-scalar theory is thus to describe possible situations of this kind. We shall give a first attempt in this direction with reference to experimental data in Sect.\ \ref{comparison}, after developing in Sects.\ \ref{scalar} and \ref{weak} the general structure of the theory and its weak-field approximation. Sect.\ \ref{conc} summarizes our conclusions.

\section{Scalar-tensor theory}
\label{scalar}

We consider a scalar-tensor theory of the BD type with an additional external scalar field $\psi $ minimally coupled to gravity with action given by (SI units are used, with $\nabla_{\mu}$ the covariant derivative) 
\begin{eqnarray}
S &=&-\frac{c^{3}}{16\pi G_{0}}\int \sqrt{-g}\phi Rd\Omega +\frac{c^{3}}{
16\pi G_{0}}\int \sqrt{-g}\frac{\omega \left( \phi \right) }{\phi }\nabla
^{\nu }\phi \nabla _{\nu }\phi d\Omega  \nonumber \\
&&+\frac{c^{3}}{16\pi G_{0}}\int \sqrt{-g}\phi \left[ \frac{1}{2}\nabla
^{\nu }\psi \nabla _{\nu }\psi -U\left( \psi \right) -J\psi \right] d\Omega +
\frac{1}{c}\int \mathcal{L}_{mat}d\Omega  \nonumber \\
&&-\frac{\varepsilon _{0}c}{4}\int \sqrt{-g}\left[ F_{\mu \nu }F^{\mu \nu
}+2\left( \nabla ^{\mu }A_{\mu }\right) ^{2}\right] d\Omega -\frac{1}{c}\int 
\sqrt{-g}j^{\nu }A_{\nu }d\Omega .  \label{SKK}
\end{eqnarray}

In order to have a non-dimensional scalar field $\phi $ of values around
unity, the constant $G_{0}$ representing Newton
gravitational constant is explicitly included, $c$ is the velocity of light in vacuum,
and $\varepsilon _{0}$ is the vacuum permittivity. $\mathcal{L}_{mat}$ is
the Lagrangian density of matter. The other symbols are conventional, $R$ is
the Ricci scalar, and $g$ the determinant of the metric tensor $g_{\mu \nu }$. The BD parameter $\omega \left( \phi \right) $ is considered a
function of $\phi $, as usually happens in the reduction to four dimensions of multidimensional theories \cite{chavineau2007}. 

The electromagnetic part in expression (\ref{SKK}) corresponds to Aharonov-Bohm (AB) action, with $F_{\mu \nu }=\nabla _{\mu}A_{\nu }-\nabla _{\nu }A_{\mu }$, given in terms of the electromagnetic
four-vector $A_{\nu }$, whose source is the four-current $j^{\nu }$. $U $ and $J$ are, respectively, the potential and source of the field $\psi $. 
The source $J$ contains contributions from the matter, electromagnetic and scalar $\phi $ fields.

In the original theory by Mbelek and Lachi\`eze-Rey (MLR) the contributions to $J$ from those fields are given in terms of the traces of their energy-momentum tensors (EMT). However, since for Maxwell action the trace of its EMT is zero, the authors chose the scalar $F_{\mu \nu }F^{\mu \nu }$ instead in the expression of $J$. The resulting expression could thus be written as
\begin{eqnarray}
J &=&\beta _{mat}\left( \psi ,\phi \right) \frac{8\pi G_{0}}{c^{4}}
T^{mat}+\beta _{EM}\left( \psi ,\phi \right) \frac{4\pi G_{0}\varepsilon _{0}
}{c^{2}}F_{\mu \nu }F^{\mu \nu }  \nonumber \\
&&+\beta _{\phi }\left( \psi ,\phi \right) T^{\phi },  \label{source}
\end{eqnarray}
where the $\beta ^{\prime }$s are non dimensional functions of the scalar
fields, $T^{mat}$ is the trace of the EMT of matter and $
T^{\phi }$ is the trace of the EMT of $\phi$ 
\[
T_{\mu \nu }^{\phi }=\nabla _{\mu }\nabla _{\nu }\phi -\nabla ^{\gamma
}\nabla _{\gamma }\phi g_{\mu \nu }+\frac{\omega \left( \phi \right) }{\phi }
\left( \nabla _{\mu }\phi \nabla _{\nu }\phi -\frac{1}{2}\nabla ^{\gamma
}\phi \nabla _{\gamma }\phi g_{\mu \nu }\right) . 
\]

An unsatisfactory feature of expression (\ref{source}) is that the resulting equations for the fields depend on $F_{\mu \nu }F^{\mu \nu }=2\left( B^{2}-E^{2}/c^{2}\right) $, so that the electric and magnetic fields enter as source terms asymmetrically. 

Another problem with this expression of $J$ is that if the theory is applied to explain some of the controversial claims mentioned in the Introduction, it is hard to justify why usual electromagnetic fields do not manifest strong gravitational effects \cite{minotti2013scalar}. 

All these difficulties can in principle be avoided considering AB electrodynamics in which the potentials are fully determined by the sources and have a direct physical effect. In this way, for instance, the electromagnetic interaction of $\psi $ can be expressed in terms of $\left( \nabla ^{\mu }A_{\mu }\right) ^{2}$, instead
of $F_{\mu \nu }F^{\mu \nu }$. In particular, since the AB scalar $S=\nabla ^{\mu }A_{\mu }$
is zero if all e.m. sources satisfy local conservation, there are no anomalous effects in ordinary situations.

Furthermore, in AB electrodynamics the trace of the electromagnetic EMT $T_{\mu \nu }^{AB}$ is $
T^{AB}=-2\varepsilon _{0}c^{2}\nabla ^{\mu }\left( SA_{\mu }\right) $, not generally zero, so that 
the interaction of the scalars with matter should include that of the electromagnetic content of that matter. 

With these considerations, we now take, instead of (\ref{source}), the following expression for the source $J$
\begin{eqnarray}
J &=&\beta _{mat}\left( \psi ,\phi \right) \frac{8\pi G_{0}}{c^{4}}\left(
T^{mat}+T^{AB}\right) +\beta _{AB}\left( \psi ,\phi \right) \frac{8\pi
G_{0}\varepsilon _{0}}{c^{2}}S^{2}  \nonumber \\
&&+\beta _{\phi }\left( \psi ,\phi \right) T^{\phi }.  \label{ABsource}
\end{eqnarray}

We proceed now to derive the equations of the presented theory. For this, the variation of the action (\ref{SKK}) is considered relative to the fields $g^{\mu \nu}$, $\phi$, $\psi$, and $A^{\mu}$. 

Variation of (\ref{SKK}) with respect to $g^{\mu \nu }$ results in the Einstein equations
\begin{eqnarray}
\phi \left( R_{\mu \nu }-\frac{1}{2}Rg_{\mu \nu }\right) &=&\frac{8\pi G_{0}
}{c^{4}}\left( T_{\mu \nu }^{AB}+T_{\mu \nu }^{mat}\right) +T_{\mu \nu
}^{\phi }  \nonumber \\
&&+\frac{\phi }{2}\left( \nabla _{\mu }\psi \nabla _{\nu }\psi -\frac{1}{2}
\nabla ^{\gamma }\psi \nabla _{\gamma }\psi g_{\mu \nu }\right)  \nonumber \\
&&+\frac{\phi }{2}\left( U+J\psi \right) g_{\mu \nu }.  \label{Glm}
\end{eqnarray}

Variation with respect to $\phi $ gives 
\begin{eqnarray*}
\phi R+2\omega \nabla ^{\nu }\nabla _{\nu }\phi &=&\left( \frac{\omega }{
\phi }-\frac{d\omega }{d\phi }\right) \nabla ^{\nu }\phi \nabla _{\nu }\phi
\\
&&-\psi \phi \frac{\partial J}{\partial \phi }+\phi \left[ \frac{1}{2}\nabla
^{\nu }\psi \nabla _{\nu }\psi -U\left( \psi \right) -J\psi \right].
\end{eqnarray*}
Using the contraction of (\ref{Glm}) with $g^{\mu
\nu }$ to replace $R$, as in the original derivation of BD theory  \cite{brans-dicke1961}, the previous expression can be rewritten as
\begin{eqnarray}
\left( 2\omega +3\right) \nabla ^{\nu }\nabla _{\nu }\phi &=&-\frac{d\omega 
}{d\phi }\nabla ^{\nu }\phi \nabla _{\nu }\phi +\frac{8\pi G_{0}}{c^{4}}
\left( T^{mat}+T^{AB}\right)  \nonumber \\
&&+\phi \left[ \frac{1}{2}\nabla ^{\nu }\psi \nabla _{\nu }\psi -U\left(
\psi \right) -J\psi \right] -\psi \phi \frac{\partial J}{\partial \phi }.
\label{phi}
\end{eqnarray}

The variation with respect to $\psi $ results in
\begin{equation}
\nabla ^{\nu }\nabla _{\nu }\psi +\frac{1}{\phi }\nabla ^{\nu }\psi \nabla
_{\nu }\phi =-\frac{\partial U}{\partial \psi }-J-\frac{\partial J}{\partial
\psi }\psi .  \label{psi}
\end{equation}

Finally, the non-homogeneous AB equations are obtained by varying (\ref{SKK}) with
respect to $A_{\nu }$,
\begin{equation}
\nabla _{\mu }F^{\mu \nu }=\mu _{0}j^{\nu }-\nabla ^{\nu }S-\nabla ^{\nu
}\left( \beta _{AB}\phi \psi S\right) .  \label{ABED}
\end{equation}
with $\mu _{0}$ the vacuum permeability.

Having included $G_{0}$, it is understood that $\phi $ takes values around its VEV $\phi _{0}=1$. The scalar $\psi $ is also dimensionless and of VEV $\psi _{0}$.

Equations (\ref{Glm}) to (\ref{ABED}) still require the explicit expressions of the $\beta$ functions. In order to proceed we will consider the weak-field limit of the theory, in which the only required information on these functions is the value of their derivatives evaluated at the VEVs $\phi_{0}$ and $\psi_{0}$. Besides, this weak-field limit should be applicable to the analysis of the experiments considered below, while containing a minimum of parameters to be determined.

\section{Weak-field approximation}
\label{weak}

We now consider the weak field approximation for values of $g_{\mu \nu }$ around $\eta
_{\mu \nu }$, taken as those of flat Minkowski space-time with signature (1,-1,-1,-1), in such a way that $g_{\mu \nu }=\eta _{\mu \nu }+h_{\mu \nu }$. 

At the lowest order in $h_{\mu \nu}$ we have
\begin{equation*}
R_{\mu \nu }-\frac{1}{2}R\eta _{\mu \nu }=\frac{1}{2}\left( -\eta ^{\gamma
\delta }\partial _{\gamma \delta }\overline{h}_{\mu \nu }+\partial _{\gamma
\mu }\overline{h}_{\nu }^{\gamma }+\partial _{\gamma \nu }\overline{h}_{\mu
}^{\gamma }-\eta _{\mu \nu }\partial _{\gamma \delta }\overline{h}^{\gamma
\delta }\right) ,
\end{equation*}
in which
\begin{equation*}
\overline{h}_{\mu \nu }\equiv h_{\mu \nu }-\frac{1}{2}h\eta _{\mu \nu },
\end{equation*}
where
\begin{equation*}
h\equiv \eta ^{\gamma \delta }h_{\gamma \delta }=-\eta ^{\gamma \delta }
\overline{h}_{\gamma \delta },
\end{equation*}
and where $\partial_{\mu \nu}$ denotes the second order derivative $\partial^{2}/\partial x^{\mu}\partial x^{\nu}$.

The system (\ref{Glm})-(\ref{ABED}) can then be written, to lowest order in $
\overline{h}_{\mu \nu }$ and in the perturbations around the VEV's of $\phi $ and $\psi $, as 
\begin{equation}
-\eta ^{\gamma \delta }\partial _{\gamma \delta }\overline{h}_{\mu \nu }=
\frac{16\pi G_{0}}{c^{4}}T_{\mu \nu }^{mat}+2\left( \partial _{\mu \nu }\phi
-\eta ^{\gamma \delta }\partial _{\gamma \delta }\phi \eta _{\mu \nu
}\right) ,  \label{Gik0}
\end{equation}
with the gauge
\begin{equation}
\partial _{\gamma }\overline{h}_{\nu }^{\gamma }=0.  \label{LG}
\end{equation}

The rest of the equations are readily found to reduce to

\begin{equation}
\left( 2\omega _{0}+3\right) \eta ^{\gamma \delta }\partial _{\gamma \delta
}\phi =\frac{8\pi G_{0}}{c^{4}}T^{mat}-\left. \frac{\partial J}{\partial
\phi }\right\vert _{\phi _{0},\psi _{0}}\phi _{0}\psi _{0},  \label{dphi0}
\end{equation}
\begin{equation}
\eta ^{\gamma \delta }\partial _{\gamma \delta }\psi =-\left. \frac{\partial
J}{\partial \psi }\right\vert _{\phi _{0},\psi _{0}}\phi _{0}\psi _{0},
\label{dpsi0}
\end{equation}
\begin{equation}
\partial _{\mu }F^{\mu \nu }=\mu _{0}j^{\nu }-\partial ^{\nu }S-\partial
^{\nu }\left( \beta S\right) ,  \label{ABE0}
\end{equation}
where $\omega _{0}=\omega \left( \phi _{0}\right) $, and
\begin{equation}
\beta=\phi _{0}\psi _{0}\left( \left. \frac{\partial \beta _{AB}}{
\partial \phi }\right\vert _{\phi _{0},\psi _{0}}\phi
+\left. \frac{\partial \beta _{AB}}{\partial \psi }\right\vert _{\phi
_{0},\psi _{0}}\psi \right) ,  \label{betanu}
\end{equation}
in which we recall that, in order not to introduce awkward notations, $\phi$ and $\psi$ represent the deviations of the scalars relative to their respective VEVs.

In these equations it was considered that in order to recover the usual physics when the scalar fields are not excited, one must have $U\left( \psi
_{0}\right) =J\left( \psi _{0},\phi _{0}\right) =0$. Also, it was taken into account that the expectation value $\psi _{0}$ satisfies $U^{\prime }\left( \psi _{0}\right)=0$.

Furthermore, due to the minimal coupling of matter to gravity considered in the action (\ref{SKK}) the motion of neutral matter correspond to geodesics of the metric. 
In the weak-field limit the geodesic equation of motion of slowly moving neutral masses can be expressed
as the action of a specific force (per unit mass) acting in flat Minkowski space-time of expression (Latin indices correspond to the spatial coordinates and $0$ to $ct$)
\begin{equation}
f_{i}=-\frac{c^{2}}{4}\frac{\partial }{\partial x_{i}}\left( \overline{h}
_{00}+\overline{h}_{kk}\right) +c\frac{\partial \overline{h}_{0i}}{\partial t
}.  \label{forcepermass}
\end{equation}

Introducing the D'Alembertian operator
\begin{equation*}
\square =\eta ^{\gamma \delta }\partial _{\gamma \delta }=\frac{1}{c^{2}}
\frac{\partial ^{2}}{\partial t^{2}}-\nabla ^{2},
\end{equation*}
applying it to the force equation (\ref{forcepermass}), and using Eq. (\ref
{Gik0}), one easily obtains
\begin{eqnarray*}
\square f_{i} &=&\frac{4\pi G_{0}}{c^{2}}\frac{\partial }{\partial x_{i}}
\left( T_{00}^{mat}+T_{kk}^{mat}\right) -\frac{16\pi G_{0}}{c^{3}}\frac{
\partial T_{0i}^{mat}}{\partial t} \\
&&+\frac{c^{2}}{2}\frac{\partial }{\partial x_{i}}\left( \square \phi -\frac{
2}{c^{2}}\frac{\partial ^{2}\phi }{\partial t^{2}}\right) .
\end{eqnarray*}

From Eqs. (\ref{dphi0}) and (\ref{dpsi0}), retaining in the EMT of matter only the dominant component $T_{00}^{mat}$, the last expression can be conveniently recast as 
\begin{equation*}
f_{i}=-\frac{\partial \chi }{\partial x_{i}},
\end{equation*}
where the "gravitational potential" $\chi $ satisfies
\begin{eqnarray}
\square \chi &=&-\frac{4\pi G_{0}}{c^{2}}T_{00}^{mat}-\frac{4\pi G_{0}}{
c^{2}\left( 2\omega _{0}+3\right) }T_{00}^{mat}  \notag \\
&&+\frac{\partial ^{2}\phi }{\partial t^{2}}+\frac{c^{2}}{2}\frac{\phi
_{0}\psi _{0}}{2\omega _{0}+3}\left. \frac{\partial J}{\partial \phi }
\right\vert _{\phi _{0},\psi _{0}}.  \label{potential}
\end{eqnarray}

The first term in this expression corresponds to Newton gravity, while the second term is the contribution of matter to the gravitational potential through the scalar $\phi $, as obtained in BD theory.

Making explicit the equation of the scalar $\phi $, Eq. (\ref{dphi0}), with
the expression of the source $J$, Eq. (\ref{ABsource}), one has for $\phi$
\begin{eqnarray}
\square \phi &=&\frac{8\pi G_{0}}{\left( 2\omega _{0}+3\right) c^{4}}\left(
1-\psi _{0}\left. \frac{\partial \beta _{mat}}{\partial \phi }\right\vert
_{\phi _{0},\psi _{0}}\right) T^{mat}  \notag \\
&&-\frac{8\pi G_{0}\varepsilon _{0}\phi _{0}\psi _{0}}{\left( 2\omega
_{0}+3\right) c^{2}}\left. \frac{\partial \beta _{AB}}{\partial \phi }
\right\vert _{\phi _{0},\psi _{0}}S^{2},  \label{lapphi}
\end{eqnarray}
where the contribution from $\phi $ itself as its source was not considered
because, even if it is present, in the weak-field approximation one has
\begin{equation*}
T^{\phi }=-3\square \phi ,
\end{equation*}
and so its effect amounts to a redefinition of the rest of the coefficients
in the equations for $\phi $ and $\psi $. 

 An interesting consequence of Eq. (\ref{lapphi}), also resulting from the original MLR theory, as remarked in \cite{minotti2013scalar}, is that if 
\begin{equation*}
\psi _{0}\left. \frac{\partial \beta _{mat}}{\partial \phi }\right\vert
_{\phi _{0},\psi _{0}}\simeq 1
\end{equation*}
the most stringent experimental bounds on the gravitational effects of the BD scalar $\phi$  could be satisfied with values of the BD parameter $\omega_{0}$ close to one, instead of larger than $4 \times 10^{4}$ \cite{bertotti2003}, thus making BD theory more satisfactory.    

Proceeding in analogous manner, the equation for $\psi $ is expressed as 
\begin{equation*}
\square \psi =-\frac{8\pi G_{0}\phi _{0}\psi _{0}}{c^{4}}\left. \frac{
\partial \beta _{mat}}{\partial \psi }\right\vert _{\phi _{0},\psi
_{0}}T^{mat}-\frac{8\pi G_{0}\varepsilon _{0}\psi _{0}}{c^{2}}\left. \frac{
\partial \beta _{AB}}{\partial \psi }\right\vert _{\phi _{0},\psi _{0}}S^{2}.
\end{equation*}

The contributions other than the matter to the potential $\chi $ can then be
obtained from (\ref{potential}) as
\begin{equation}
\square \chi=\frac{\partial ^{2}\phi }{\partial t^{2}}+\frac{4\pi
G_{0}\varepsilon _{0}\phi _{0}\psi _{0}}{2\omega _{0}+3}\left. \frac{
\partial \beta _{AB}}{\partial \phi }\right\vert _{\phi _{0},\psi _{0}}S^{2}.
\label{potem}
\end{equation}

Finally, the equation for $A^{\nu }$ is Eq. (\ref{ABE0}) explicitly written as 
\begin{equation}
\square A^{\nu }=\mu _{0}j^{\nu }-\partial ^{\nu }\left( \beta S\right) .  \label{LapAnu}
\end{equation}

In this way, given the source $j^{\nu }$ we can determine its gravitational
effect through Eqs. (\ref{potem}), (\ref{LapAnu}), (\ref{betanu}) and 
\begin{subequations}
\label{DalamPhiPsi}
\begin{eqnarray}
\square \phi &=&-\frac{8\pi G_{0}\varepsilon _{0}\phi _{0}\psi _{0}}{\left(
2\omega _{0}+3\right) c^{2}}\left. \frac{\partial \beta _{AB}}{\partial \phi 
}\right\vert _{\phi _{0},\psi _{0}}S^{2}, \\
\square \psi &=&-\frac{8\pi G_{0}\varepsilon _{0}\psi _{0}}{c^{2}}\left. 
\frac{\partial \beta _{AB}}{\partial \psi }\right\vert _{\phi _{0},\psi
_{0}}S^{2}.
\end{eqnarray}
\end{subequations}

\section{Special solutions and comparison with possibly related experiments}
\label{comparison}

In this section we compare the results in the previous sections with two
series of published experiments whose theoretical basis could be discussed
within the present model. One is that of pulsed electric discharges on
high-temperature superconductors, resulting in traveling pulses with
apparently gravitational effects (\cite{gallerati2022interaction} and refs.). The other is related to the possibility of anomalous forces in closed microwave cavities (\cite{neunzig2022} and refs.).

\subsection{1-D traveling solutions}

If traveling pulses causing a force associated to the non-matter gravitational potential $\chi$ exist in the theory, they must obey the field equations found above. We thus proceed to obtain traveling solutions of those equations that could be associated to experiments using transient electric discharges \cite{modanese2013APR}. 
  
In order to proceed, it is convenient to define the constants 
\begin{subequations}
\label{CKs}
\begin{eqnarray}
K_{\phi } &=&\frac{8\pi G_{0}\varepsilon _{0}}{\left( 2\omega _{0}+3\right)
c^{2}}C_{\phi }, \\
K_{\psi } &=&\frac{8\pi G_{0}\varepsilon _{0}}{c^{2}}C_{\psi }, \\
C_{\phi } &=&\phi _{0}\psi _{0}\left. \frac{\partial \beta _{AB}}{\partial
\phi }\right\vert _{\phi _{0},\psi _{0}}, \\
C_{\psi } &=&\phi _{0}\psi _{0}\left. \frac{\partial \beta _{AB}}{\partial
\psi }\right\vert _{\phi _{0},\psi _{0}}.
\end{eqnarray}
\end{subequations}

Taking into account that $8\pi G_{0}\varepsilon _{0}c^{-2}\simeq 1.6\times
10^{-37}$A$^{2}$N$^{-2}$, the perturbations of the scalars about their VEV,
as given by Eqs. (\ref{DalamPhiPsi}), are expected to be very small.
Consequently, the second term in the right-hand side of Eq. (\ref{LapAnu})
can be neglected as compared with the material sources of the four-potential 
$A^{\nu }$. Taking the four-divergence of the resulting equation we obtain
for the AB\ scalar $S=\partial _{\nu }A^{\nu }$ the equation 
\begin{equation}
\square S=\mu _{0}\partial _{\nu }j^{\nu },  \label{DAlamS}
\end{equation}
which is the usual equation studied in AB electrodynamics, valid in flat
Minkowski space-time. 

The solutions of this equation are thus determined entirely by the material
sources. Once $S$ is obtained, the scalars and the gravitational potential
can be determined using Eqs. (\ref{potem}) and (\ref{DalamPhiPsi}).

As a first example we study one-dimensional solutions depending on time and
on the spatial Cartesian coordinate $x$. We consider that at $t\geq 0$ the
material sources generate a given $S\left( x,t\right) $ satisfying outside
those sources the corresponding expression of Eq. (\ref{DAlamS})
\begin{equation*}
c^{-2}\partial _{tt}S-\partial _{xx}S=0,
\end{equation*}
so that the scalars and potential $\chi $\ are determined by the equations 
\begin{eqnarray*}
c^{-2}\partial _{tt}\phi -\partial _{xx}\phi &=&-K_{\phi }S^{2}, \\
c^{-2}\partial _{tt}\psi -\partial _{xx}\psi &=&-K_{\psi }S^{2}, \\
c^{-2}\partial _{tt}\chi -\partial _{xx}\chi &=&\partial _{tt}\phi +\frac{
K_{\phi }c^{2}}{2}S^{2}.
\end{eqnarray*}

We can thus obtain the potentials from the solution of the general equation
\begin{equation}
c^{-2}\partial _{tt}u-\partial _{xx}u=g\left( x,t\right) ,  \label{wave1dnh}
\end{equation}
with prescribed $g\left( x,t\right) $, and initial conditions $u\left(
x,0\right) =\partial _{t}u\left( x,0\right) =0$.

The general solution of this problem is well known, given by
\begin{equation}
u\left( x,t\right) =c\int_{0}^{t}v\left( x,t-s;s\right) ds,  \label{usolgen}
\end{equation}
where
\begin{equation}
v\left( x,t;s\right) =\frac{1}{2}\int_{x-ct}^{x+ct}g\left( y,s\right) dy.
\label{vsolgen}
\end{equation}

As a simple, but relevant example that allows for a fully analytical
solution, we consider that at $t=0$ and $x=0$ the material sources generate
a very localized pulse of $S$ that travels in the positive $x$\ direction,
so that for its model we take, for $t>0$,
\begin{equation*}
S^{2}=S_{0}^{2}\delta \left[ \alpha \left( x-ct\right) \right] ,
\end{equation*}
with $S_{0}$ and $\alpha $ constants with units of $S$ and of inverse
length, respectively, with $\alpha ^{-1}$ a measure of the pulse
"spatial-width".

Using relations (\ref{usolgen}) and (\ref{vsolgen}) the solution for the
scalars is readily obtained as $\phi \left( x,t\right) =-K_{\phi }h\left(
x,t\right) $, and $\psi \left( x,t\right) =-K_{\psi }h\left( x,t\right) $,
where 
\begin{equation}
h\left( x,t\right) =\frac{S_{0}^{2}}{4\alpha }\left( x+ct\right) \left\{ H
\left[ \alpha \left( x+ct\right) \right] -H\left[ \alpha \left( x-ct\right) 
\right] \right\} ,  \label{hxt}
\end{equation}
with $H$ Heaveside unit-step function.

As for the potential $\chi $ we have for the obtained solution of $\phi $
\begin{equation*}
\partial _{tt}\phi =-\frac{K_{\phi }S_{0}^{2}c^{2}}{2}\left\{ \delta \left[
\alpha \left( x-ct\right) \right] +\delta \left[ \alpha \left( x+ct\right) 
\right] \right\} ,
\end{equation*}
so that the equation for $\chi $ is of the general form (\ref{wave1dnh}) with
\begin{equation*}
g\left( x,t\right) =-\frac{K_{\phi }S_{0}^{2}c^{2}}{2}\delta \left[ \alpha
\left( x+ct\right) \right] ,
\end{equation*}
whose solution is
\begin{equation*}
\chi \left( x,t\right) =-\frac{K_{\phi }S_{0}^{2}c^{2}}{8\alpha }\left(
x-ct\right) \left\{ H\left[ \alpha \left( x-ct\right) \right] -H\left[
\alpha \left( x+ct\right) \right] \right\} .
\end{equation*}

The corresponding specific force $f_{x}=-\partial _{x}\chi $ is
\begin{equation}
f_{x}=\frac{K_{\phi }S_{0}^{2}c^{2}}{8\alpha }\left\{ H\left[ \alpha \left(
x-ct\right) \right] -H\left[ \alpha \left( x+ct\right) \right] -2\alpha
x\delta \left[ \alpha \left( x+ct\right) \right] \right\} .
\label{force1dgen}
\end{equation}

It is interesting to note that a massive object placed at $x>0$ suffers a
permanent force for all $t>x/c$, while on objects placed at $x<0$ a similar
constant force acts\ for all $t>-x/c$, but with the addition of an opposite
impulsive force, acting at $t=-x/c$, thus reducing the total impulse gained
by bodies at $x<0$, relative to that gained by those at $x>0$.

Also, for $K_{\phi }>0$ ($K_{\phi }<0$) the permanent part of the force acts
in the opposite (same) direction of the pulse motion. The impulsive
component acting on objects at $x<0$ is always opposite to the permanent
part, and its magnitude grows with the distance from $x=0$.

These characteristics of the pulse can in principle be tested in the experiments of the type mentioned in \cite{modanese2013APR}.

As a second example we consider a one-dimensional, spherically symmetric
case depending on radius $r$, and in which the solution of (\ref{DAlamS})
outside its sources is of the general form 
\begin{equation*}
S\left( r,t\right) =\frac{f\left( r-ct\right) }{r}.
\end{equation*}

The general solution for the scalar $\phi $ in the case of no boundaries is given as 
\begin{equation}
\phi \left( \mathbf{x},t\right) =-\frac{K_{\phi }}{4\pi }\int \frac{
S^{2}\left( \mathbf{x}^{\prime },t^{\prime }\right) }{\left\vert \mathbf{x}
^{\prime }-\mathbf{x}\right\vert }d^{3}x^{\prime },  \label{phigeneral}
\end{equation}
where
\begin{equation}
t^{\prime }=t-\frac{\left\vert \mathbf{x}^{\prime }-\mathbf{x}\right\vert }{c
}.  \label{tprime}
\end{equation}

In the spherically symmetric case considered we have
\begin{equation*}
\phi \left( r,t\right) =-\frac{K_{\phi }}{2}\int_{0}^{ct}dr^{\prime
}\int_{-1}^{1}d\mu \frac{f^{2}\left( r^{\prime }-ct+\sqrt{r^{\prime
2}+r^{2}-2\mu rr^{\prime }}\right) }{\sqrt{r^{\prime 2}+r^{2}-2\mu
rr^{\prime }}}.
\end{equation*}

We consider again a very narrow $S$ pulse, in order to model the function $f$ as
\begin{equation*}
f\left( \xi \right) =\alpha ^{-2}S_{0}^{2}\delta \left( \xi \right) ,
\end{equation*}
with, as before, $S_{0}$ and $\alpha $ constants with units of $S$ and of inverse length, respectively. 

An elementary integration gives the solution,
valid for $r\leq ct$,
\begin{equation*}
\phi \left( r,t\right) =\frac{K_{\phi }\alpha ^{-2}S_{0}^{2}}{2r}\ln \left( 
\frac{ct-r}{ct+r}\right) .
\end{equation*}

The equation for $\chi $ is thus
\begin{equation*}
\square \chi =-\frac{2K_{\phi }\alpha ^{-2}S_{0}^{2}c^{3}t}{\left(
c^{2}t^{2}-r^{2}\right) ^{2}}+\frac{K_{\phi }\alpha ^{-2}S_{0}^{2}c^{2}}{
2r^{2}}\delta \left( r-ct\right) .
\end{equation*}

If we now define the function
\begin{equation*}
g\left( r,t\right) =\frac{K_{\phi }\alpha ^{-2}S_{0}^{2}c^{3}t}{2\left(
c^{2}t^{2}-r^{2}\right) },
\end{equation*}
and note that
\begin{equation*}
\square g=-\frac{2K_{\phi }\alpha ^{-2}S_{0}^{2}c^{3}t}{\left(
c^{2}t^{2}-r^{2}\right) ^{2}},
\end{equation*}
we see that the function $\chi -g$ satisfies the initial condition of being
zero at $t=0$ and also satisfies the wave equation with a delta source, as
does $\phi $, with similar initial conditions. In this way, the solution for $
\chi -g$ is obtained as that of $\phi $, and results in $\chi $ given by 
\begin{equation*}
\chi \left( r,t\right) =\frac{K_{\phi }\alpha ^{-2}S_{0}^{2}c^{3}t}{2\left(
c^{2}t^{2}-r^{2}\right) }-\frac{K_{\phi }\alpha ^{-2}S_{0}^{2}c^{2}}{2r}\ln
\left( \frac{ct-r}{ct+r}\right) .
\end{equation*}

The corresponding specific radial force is
\begin{equation*}
f_{r}=-\partial _{r}\chi =-\frac{K_{\phi }\alpha ^{-2}S_{0}^{2}c^{2}}{2r^{2}}
\left[ \ln \left( \frac{ct-r}{ct+r}\right) +\frac{2rc^{3}t^{3}}{\left(
c^{2}t^{2}-r^{2}\right) ^{2}}\right] .
\end{equation*}

\subsection{Soliton solution}

As another example\ in which the smallness of $K_{\phi }$\ and $K_{\psi }$
is not assumed from the beginning, we consider a one-dimensional traveling
wave solution with fixed shape, or soliton solution, of the form 
\begin{eqnarray*}
A^{\nu } &=&\left( 0,f\left( x-vt\right) ,0,0\right) , \\
\phi &=&\phi\left( x-vt\right) , \\
\psi &=&\psi\left( x-vt\right) ,
\end{eqnarray*}
where the velocity $v$ is in general different from $c$.

The chosen electromagnetic potential corresponds to a scalar wave with
scalar and longitudinal electric fields given by (the prime denotes
derivative of the function with respect to its argument $\xi=x-vt$)
\begin{eqnarray*}
S &=&f^{\prime }\left(\xi\right) , \\
E_{x} &=&vf^{\prime }\left(\xi\right) =vS.
\end{eqnarray*}

We thus have, outside the sources, from Eqs. (\ref{betanu}), (\ref{LapAnu}) and (\ref{DalamPhiPsi}), where now the $S\beta$ term in Eq. (\ref{LapAnu}) is not neglected,
\begin{eqnarray*}
\left( \frac{v^{2}}{c^{2}}-1\right) S^{\prime } &=&\left( \beta S\right)
^{\prime }, \\
\left( \frac{v^{2}}{c^{2}}-1\right) \beta ^{\prime \prime } &=&-\Lambda
S^{2},
\end{eqnarray*}
with
\begin{equation}
\Lambda =C_{\phi }K_{\phi }+C_{\psi }K_{\psi },  \label{lambda}
\end{equation}
which, by its definition, is positive definite.

These equations are readily reduced to the single equation for $\Sigma =S^{-1}$
\begin{equation}
\Sigma ^{2}\Sigma ^{\prime \prime }=-\kappa ,  \label{eqSigma}
\end{equation}
with $\kappa $ an arbitrary constant.

Using that
\begin{equation*}
\Sigma ^{\prime \prime }=\frac{d\Sigma ^{\prime }}{d\xi }=\frac{d\Sigma
^{\prime }}{d\Sigma }\Sigma ^{\prime }
\end{equation*}
we can easily obtain the general solution of Eq. (\ref{eqSigma}) in implicit
form, given in terms of $S$,  
\begin{equation*}
\int \frac{dS}{S^{2}\sqrt{c_{1}+2\kappa S}}=\pm \left( \xi +c_{2}\right) ,
\end{equation*}
with $c_{1,2}$ arbitrary constants. The indefinite integral is simple and
gives
\begin{equation}
\pm \text{arctanh}\sqrt{1\pm u}-\frac{\sqrt{1\pm u}}{u}=\pm \frac{c_{1}^{3/2}
}{2\kappa }\left( \xi -\xi _{0}\right) ,  \label{Simpli}
\end{equation}
where we have written $c_{2}=-\xi _{0}$, and 
\begin{equation*}
u=\frac{2\kappa }{c_{1}}S.
\end{equation*}

Unfortunately, relation (\ref{Simpli}) cannot be inverted explicitly, but it
can be easily plotted to verify that $S\left( \xi \right) $ is a localized
function with maximum at $\xi _{0}$, of value $c_{1}/\left( 2\kappa \right) $, with a width $2\kappa c_{1}^{-3/2}$, and which goes to zero for $\xi
\rightarrow \pm \infty $. 

\subsection{Resonant electromagnetic cavities}
\label{resonant}

We study in this section the gravitational effects that could be expected,
according to the previous theory, in resonant electromagnetic closed
cavities. The condition of resonance is considered in order that
sufficiently high intensity currents develop in the cavity internal surface,
so that if a fraction of those currents happens to be not conserved locally,
observable effects could be expected. The model to account for the
non-conserved part of the current will be that developed in \cite{EPJC2023}, the $\gamma $ model, and we consider in particular that the material media
present are not polarizable.

We consider harmonic in time solutions for the resonant mode in the cavity,
described by Maxwell theory, with corresponding conserved current and charge
densities present in the internal cavity walls. The $\gamma $ model allows
to describe in terms of those magnitudes the effect of a small not locally
conserved current that would generate a scalar field $S$ inside the cavity,
serving as a source of the gravitational potential, as given by relation (\ref{potem}).

We further time-average all magnitudes over the rapid oscillation of the
electromagnetic fields, typically in the microwave range. Denoting this time
average by $\left\langle ...\right\rangle $, since $\left\langle \frac{
\partial }{\partial t}...\right\rangle =0$, averaging Eq. (\ref{potem}) we
obtain 
\begin{equation}
\nabla ^{2}\left\langle \chi \right\rangle =-K_{\phi }c^{2}\left\langle
S^{2}\right\rangle ,  \label{avgKi}
\end{equation}
where the definition of $K_{\phi }$ in (\ref{CKs}) was used.

The equation for $S=\partial _{\nu }A^{\nu }$ is obtained by taking the
four-divergence of Eq. (\ref{LapAnu}):
\begin{equation*}
\square S=\mu _{0}I-\square\left( \beta S\right) ,
\end{equation*}
where $I=\partial _{\nu }j^{\nu }$ is the "extra-current", which is
different from zero only if the current is not conserved locally. Using (\ref
{DalamPhiPsi}) and (\ref{CKs}) we can write
\begin{eqnarray*}
\square\beta  &=&-\left( C_{\phi }K_{\phi }+C_{\psi }K_{\psi} \right) S^{2},
\end{eqnarray*}
which indicates that $\beta S$ is non linear in the small quantities $S$ and its
derivatives. In this way, the linearized equation for $S$ is
\begin{equation*}
\square S=\mu _{0}I,
\end{equation*}
whose general solution is 
\begin{equation*}
S\left( \mathbf{x},t\right) =\frac{\mu _{0}}{4\pi }\int \frac{I\left( 
\mathbf{x}^{\prime },t^{\prime }\right) }{\left\vert \mathbf{x}-\mathbf{x}
^{\prime }\right\vert }d^{3}x^{\prime },
\end{equation*}
where $t^{\prime }=t-\left\vert \mathbf{x}-\mathbf{x}^{\prime }\right\vert /c
$. 

Using now the $\gamma $ model, we have $I=-\gamma \nabla \cdot \mathbf{j}$,
\cite{EPJC2023} where $\mathbf{j}$ is the conserved current density, and $\gamma$ a dimensionless small constant. In this way
\begin{equation*}
S\left( \mathbf{x},t\right) =-\frac{\mu _{0}\gamma }{4\pi }\int \frac{\nabla
^{\prime }\cdot \mathbf{j}^{\prime }}{\left\vert \mathbf{x}-\mathbf{x}
^{\prime }\right\vert }d^{3}x^{\prime }=\frac{\mu _{0}\gamma }{4\pi }\int 
\frac{\partial \rho ^{\prime }/\partial t^{\prime }}{\left\vert \mathbf{x}-
\mathbf{x}^{\prime }\right\vert }d^{3}x^{\prime },
\end{equation*}
where in the last equality the condition of current conservation was used to
express the solution in terms of the time derivative of the charge density $
\rho $. Noting further that the solution for the scalar electromagnetic potential $\varphi $
is given by
\begin{equation}
\varphi \left( \mathbf{x},t\right) =\frac{1}{4\pi \varepsilon _{0}}\int 
\frac{\rho \left( \mathbf{x}^{\prime },t^{\prime }\right) }{\left\vert 
\mathbf{x}-\mathbf{x}^{\prime }\right\vert }d^{3}x^{\prime },
\label{scalarpot}
\end{equation}
we can write in general that
\begin{equation}
S=\frac{\gamma }{c^{2}}\frac{\partial \varphi }{\partial t}.  \label{Sdphidt}
\end{equation}

Since we consider a powered closed cavity we have a varying potential in its
interior, and a constant potential in its exterior. This is consistent with
the results in \cite{minotti2023aharonov} showing that the scalar $S$ cannot propagate
across a normal medium, so that outside the cavity it must be $S=0$.

In this way, we can evaluate the time-averaged gravitational forces acting
on a powered closed cavity solving equations (\ref{avgKi}) and (\ref{Sdphidt}), in which the potential is evaluated using (\ref{scalarpot}) with the charge density obtained from the standard Maxwell solution for the electromagnetic fields in the cavity. 

Since the surface charge density $\sigma $ at the inside wall of the cavity
can be written in terms of the normal component of the electric field as $
\sigma =\varepsilon _{0}\mathbf{E}\cdot \mathbf{n}$, with $\mathbf{n}$ the
external unit vector at the interior wall, we can explicitly write
\begin{equation}
S=\frac{\gamma }{4\pi c^{2}}\frac{\partial }{\partial t}\oint_{\Sigma }\frac{
\mathbf{E}^{\prime }\cdot \mathbf{n}^{\prime }}{\left\vert \mathbf{x}-
\mathbf{x}^{\prime }\right\vert }d\Sigma ^{\prime },  \label{Scavity}
\end{equation}
where the integral is extended to the internal surface of the cavity $\Sigma$. 

Note in particular that those modes in which the electric field is
transverse to the cavity walls do not generate a scalar field in this model.

As a relatively simple example we consider now a cylindrical cavity of
radius $a$ and height $h$. We use cylindrical coordinates ($r$, $z$, $\theta 
$), where $z$ is the coordinate along the symmetry axis of the cavity, with
origin at the circular base of the cylinder. The fundamental mode is the TM$
_{010}$, whose electric field has only a $z$ component with expression
\begin{equation*}
E_{z}=E_{0}J_{0}\left( \lambda_{0} \frac{r}{a}\right) \cos \omega t,
\end{equation*}
in which $E_{0}$ is the electric field amplitude at the cavity axis, $J_{0}$
is the Bessel function of the first kind of order $0$, $\lambda_{0} $ its first
zero, $\lambda_{0} \simeq 2.405$, and the angular frequency of the mode is $
\omega =\lambda_{0} c/a$.

Since $\mathbf{E}\cdot \mathbf{n}=0$ at the lateral walls, only the circular
surfaces at $z=0$ and $z=h$ contribute in expression (\ref{Scavity}), which
is explicitly written as
\begin{eqnarray*}
S(r,z,\theta ,t) &=&\frac{\gamma E_{0}}{4\pi c^{2}}\frac{\partial }{\partial
t}\int_{0}^{a}\int_{0}^{2\pi }J_{0}\left( \lambda_{0} \frac{r^{\prime }}{a}
\right) \times  \\
&&\left[ \left. \frac{\cos \left( \omega t-kR\right) }{R}\right\vert
_{z=0}-\left. \frac{\cos \left( \omega t-kR\right) }{R}\right\vert _{z=h}
\right] r^{\prime }d\theta ^{\prime }dr^{\prime },
\end{eqnarray*}
where $k=\omega /c$, and
\begin{equation*}
R=\sqrt{r^{2}+r^{\prime 2}+\left( z-z^{\prime }\right) ^{2}-2rr^{\prime
}\cos \left( \theta -\theta ^{\prime }\right) }.
\end{equation*}

The expression for $S$ can be further elaborated using that $\cos \left(
\omega t-kR\right) =\cos \omega t\cos kR+\sin \omega t\sin kR$ to obtain
\begin{equation*}
S=S_{s}\cos \omega t-S_{c}\sin \omega t,
\end{equation*}
where
\begin{eqnarray*}
S_{s} &=&\frac{\gamma E_{0}\omega }{4\pi c^{2}}\int_{0}^{a}\int_{0}^{2\pi
}J_{0}\left( \lambda_{0} \frac{r^{\prime }}{a}\right) \left[ \left. \frac{\sin kR
}{R}\right\vert _{z=0}-\left. \frac{\sin kR}{R}\right\vert _{z=h}\right]
r^{\prime }d\theta ^{\prime }dr^{\prime }, \\
S_{c} &=&\frac{\gamma E_{0}\omega }{4\pi c^{2}}\int_{0}^{a}\int_{0}^{2\pi
}J_{0}\left( \lambda_{0} \frac{r^{\prime }}{a}\right) \left[ \left. \frac{\cos kR
}{R}\right\vert _{z=0}-\left. \frac{\cos kR}{R}\right\vert _{z=h}\right]
r^{\prime }d\theta ^{\prime }dr^{\prime }.
\end{eqnarray*}

We thus have
\begin{equation*}
\left\langle S^{2}\right\rangle =\frac{1}{2}\left(
S_{s}^{2}+S_{c}^{2}\right) .
\end{equation*}

The integrations to obtain $S_{s,c}$ have to be done numerically, but we can
determine an estimation of the expected values by expressing the ($r$, $z$) variables in units of $a$ to write
\begin{equation}
\left\langle S^{2}\right\rangle =\left( \frac{\gamma E_{0}\omega a}{4\pi
c^{2}}\right) ^{2}\eta ,  \label{S2estimated}
\end{equation}
where $\eta \left( \xi ,\zeta \right) $ is a non-dimensional function of the non-dimensional coordinates $\xi =r/a$ and $\zeta =z/a$.

The actual force on the cavity is obtained as
\begin{equation*}
\mathbf{F}=-\oint_{\Sigma }\nabla \left\langle \chi \right\rangle \sigma
_{m}d\Sigma ,
\end{equation*}
where $\sigma _{m}$ is the wall mass density per unit of surface (it is assumed
that the cavity walls are sufficiently thin for this approximation to
apply), and the integral is extended to the surface $\Sigma $ of the cavity.

For the axially symmetric mode considered the net force on the lateral walls
is zero if $\sigma _{m}$ is also axially symmetric. The contributions from
the base and top surfaces also cancel if $\sigma _{m}$ is the same for both
of them. A net force can thus be obtained only if one of those surfaces is
thicker than the other so that its $\sigma _{m}$ is larger. We thus consider a cylindrical cavity with asymmetric mass distribution, in which the base and lateral surfaces have a much smaller mass density than the top surface.

The force at the top surface, with uniform $\sigma _{m}$ has only a $z$
component:
\begin{equation*}
F_{z}=-2\pi \sigma _{m}\int_{0}^{a}\left. \frac{\partial \left\langle \chi
\right\rangle }{\partial z}\right\vert _{z=h}rdr.
\end{equation*}
We can estimate its value integrating Eq. (\ref{avgKi}) over the volume of the
upper half of the cavity, $V_{uh}$, to obtain
\begin{equation*}
\int_{V_{uh}}\nabla ^{2}\left\langle \chi \right\rangle d^{3}x=\oint_{\Sigma
\left( V_{uh}\right) }\nabla \left\langle \chi \right\rangle \cdot \mathbf{n}
d\Sigma =-K_{\phi }c^{2}\int_{V_{uh}}\left\langle S^{2}\right\rangle d^{3}x.
\end{equation*}
Since $\partial \left\langle \chi \right\rangle /$ $\partial z=0$ by
symmetry at the midplane of the cavity we have
\begin{equation*}
\oint_{\Sigma \left( V_{uh}\right) }\nabla \left\langle \chi \right\rangle
\cdot \mathbf{n}d\Sigma =2\pi \int_{0}^{a}\left. \frac{\partial \left\langle
\chi \right\rangle }{\partial z}\right\vert _{z=h}rdr+2\pi
a\int_{h/2}^{h}\left. \frac{\partial \left\langle \chi \right\rangle }{
\partial r}\right\vert _{r=a}dz.
\end{equation*}

The numerical result for $\left\langle S^{2}\right\rangle $ shows that its
magnitude is concentrated around the cavity axis and in the immediate
vicinity of the base and top surfaces. This indicates that $\partial
\left\langle \chi \right\rangle /$ $\partial z$ at the top surface is dominant over $\partial
\left\langle \chi \right\rangle /$ $\partial r$ evaluated at the lateral surface. In this way, we can approximate
\begin{equation*}
F_{z}\simeq -\sigma _{m}\oint_{\Sigma \left( V_{uh}\right) }\nabla
\left\langle \chi \right\rangle \cdot \mathbf{n}d\Sigma =\sigma _{m}K_{\phi
}c^{2}\int_{V_{uh}}\left\langle S^{2}\right\rangle d^{3}x,
\end{equation*}
which using (\ref{S2estimated}) gives for the force estimation
\begin{eqnarray*}
F_{z} &\simeq &\sigma _{m}K_{\phi }c^{2}\left( \frac{\gamma E_{0}\omega a}{
4\pi c^{2}}\right) ^{2}2\pi a^{3}\int \eta \xi d\xi d\zeta  \\
&=&\sigma _{m}K_{\phi }\frac{\gamma ^{2}E_{0}^{2}\lambda_{0} ^{2}}{8\pi }
a^{3}\int \eta \xi d\xi d\zeta .
\end{eqnarray*}

The integral in this expression can be easily done
numerically and has a value of order unity. For instance, for $h=a$ the integral has a value of approximately $3.6$, while if $h=3a$ its value is
around $9.2$. 

To more practically determine the magnitudes involved we can use the definition of the quality factor of the cavity
\begin{equation*}
Q_{cav}=\frac{\omega \left\langle U\right\rangle }{\left\langle
W\right\rangle },
\end{equation*}
where $\left\langle U\right\rangle $ is the time averaged electromagnetic energy inside the cavity, and $\left\langle W\right\rangle $ the mean
dissipated power. For the TM$_{010}$ mode considered we have
\begin{equation*}
\left\langle U\right\rangle =\varepsilon _{0}E_{0}^{2}\pi
h\int_{0}^{a}J_{0}^{2}\left( \lambda_{0} \frac{r}{a}\right) rdr\simeq
0.42\varepsilon _{0}E_{0}^{2}a^{2}h.
\end{equation*}

On the other hand, the permanent regime of the powered cavity is reached
when the mean power fed to the cavity, $P$, equals the mean dissipated power 
$\left\langle W\right\rangle $, so that $\left\langle U\right\rangle
=Q_{cav}P/\omega $, which allows us to write for the force
\begin{equation}
F_{z}\simeq 0.23\gamma ^{2}\frac{\sigma _{m}K_{\phi }Q_{cav}Pa^{2}}{
c\varepsilon _{0}h}\int \eta \xi d\xi d\zeta \simeq \gamma ^{2}\frac{\sigma
_{m}K_{\phi }Q_{cav}Pa^{2}}{c\varepsilon _{0}h},  \label{Fzestimated}
\end{equation}
where the relation $\omega =\lambda_{0} c/a$ was used, and in the right-most
expression the non-dimensional factor of the order of unity was left out.

If we further note that the mass on which this force acts is of the order $\sigma_{m}a^{2}$, the mean specific force  can be estimated as 

\begin{equation}
f_{z} \simeq \gamma ^{2}\frac{K_{\phi }Q_{cav}P}{c\varepsilon _{0}h},.  \label{fzcav}
\end{equation}

\subsection{Magnitudes estimation}
\label{MO}

In order to relate the obtained solutions to possible experiments we need some estimation of the expected values of the free parameters of the theory in its weak-field limit. To estimate those magnitudes we consider the following. By explicit use
of Newton gravitational constant in the action of the theory, the VEV of $\phi $ is taken as 1. Also, the explicit use of the fundamental constants
in the source term $J$, in order to work with non-dimensional functions $\beta $, should further impose that the order of those functions is also one.

On the other hand, we have in principle no way to ascertain the magnitude of
the VEV of the scalar $\psi $, so that its value $\psi _{0}$ is thus far
left to be determined experimentally.

In order to proceed we make the following hypothesis on the magnitudes of
the derivatives of $\beta _{AB}\left( \phi ,\psi \right) $ 
\begin{equation*}
\frac{\partial \ln \beta _{AB}}{\partial \ln \phi }\sim \frac{\partial \ln
\beta _{AB}}{\partial \ln \psi }\sim 1,
\end{equation*}

With all the above considerations we have
\begin{eqnarray*}
C_{\phi } &=&\phi _{0}\psi _{0}\left. \frac{\partial \beta _{AB}}{\partial
\phi }\right\vert _{\phi _{0},\psi _{0}}\sim \psi _{0}, \\
C_{\psi } &=&\phi _{0}\psi _{0}\left. \frac{\partial \beta _{AB}}{\partial
\psi }\right\vert _{\phi _{0},\psi _{0}}\sim 1.
\end{eqnarray*}

In this way, for the specific force (\ref{force1dgen}), defining the
pulse duration as $\tau =\left( \alpha c\right) ^{-1}$, we have, as an order
of magnitude estimation, in SI units,
\begin{equation*}
f_{x}\sim K_{\phi }S_{0}^{2}c^{3}\tau \sim 10^{-29}\psi _{0}E_{0}^{2}\tau ,
\end{equation*}
where $E_{0}=cS_{0}$ is the magnitude of the longitudinal electric field of
the pulse. 
We can have an estimation of $E_{0}$ considering  that, in a transient electrical discharge in which these pulses are apparently generated, a possible component of the current that is not conserved locally gives rise to the fields $S$ and $E_{0}$ that are finally emitted into vacuum. By continuity at the material electrode surface $E_{0}$ is expected to be a fraction of the local electric field.
The reason being that, of both sources of the electric field (the electric charge and $S$) the $S$ component is continuous across the interface surface-vacuum. In this way, an efficient generation of $S$, that implies a non-negligible fraction of locally non-conserved current, corresponds to an $E_{0}$ that is also a non-negligible fraction of the normal component longitudinal electric field. The magnitude of the latter can be estimated through the buildup of the surface distribution of charge $\sigma $ at the interface, as $E_{s}\sim
\sigma /\varepsilon _{0}\sim J\tau /\varepsilon _{0}$, with $J$ the current density. 

Using now the $\gamma$-model developed in \cite{EPJC2023} to relate the non-conserved current to the conserved one, we can write that $E_{0}$ is a fraction $\gamma$ of $E_{s}$, where $\gamma$ is roughly the ratio of non-conserved to conserved current, times the fraction of the length traversed by the current along which the non-conservation applies. We can thus write
\begin{equation}
f_{x}\sim 10^{-7}\gamma^{2}\psi _{0}J^{2}\tau ^{3}.
\label{force1}
\end{equation}

In this way, an $f_{x}$ of the order of the gravitational acceleration on Earth
requires that $\gamma^{2}\psi _{0}J^{2}\tau ^{3}\sim 10^{8}$. Current densities of the
order of $10^{12}$ A m$^{-2}$ are standard in the cathode spots of vacuum arc discharges \cite{mesyats2013}, so that conditions exist for an actual experiment in which relatively strong gravitational forces could be present.

 An alternative and more conservative criterion for defining the magnitude order of $S_0$ is the following one, based on energy considerations. It is straightforward to prove that in AB electrodynamics the energy carried by an $S$-wave is equal to that of a standard e.m.\ wave in which $E=cS$. If we admit the existence of a laboratory source of $S$, we should reasonably assume that the $S$-wave generated cannot transport more energy than an equivalent strong $E$-wave in which $E=cS$. Otherwise, not only would we need an extraordinarily efficient $S$-generator, but also an extraordinarily powerful energy source. Now, take for example a radio emitter able to generate a power of 10 kW over a cross-section of 1 m$^2$. This requires a squared field $E^2\sim 10^7$ in SI units, corresponding to $S^2\sim 10^{-10}$. According to this ''energy criterion'', the specific force $f_x$ estimated from \eqref{force1dgen} is expected to be definitely smaller than found in \eqref{force1}.

 For the case of the force on resonant cavities we can similarly consider the estimation of the specific force (\ref{fzcav}). Using the above estimations for the constants of the theory an specific force of the order of the gravitational acceleration on Earth requires that, in SI units,
\begin{equation*}
 \gamma ^{2}\psi _{0} Q_{cav} P h^{-1}\sim 10^{35}.  \end{equation*}
 
\section{Conclusions}
\label{conc}

In this work a tensor-scalar theory of gravity has been introduced, that includes a novel coupling between the gravitational scalars $\phi$, $\psi$ and the scalar field $S$ of Aharonov-Bohm extended electrodynamics. After developing the formal structure of the model and writing the field equations in the weak-field approximation, we have applied them to the analysis of two kind of experiments. 

The first kind of experiments concerns impulsive anomalous forces (with duration $\tau$ of the order of $10^{-4}$ s or less) generated by high-voltage electrical discharges with superconducting electrodes. The experimental data indicate that the mechanical impulse per unit mass transferred by these forces is of the order of 1 m/s. According to our theory, a scalar e.m. wave of reasonable intensity, if existing (since it requires a source with violation of local charge conservation) is actually coupled to the scalar gravitational components and generates a force depending on a few parameters which are difficult to measure. We have identified two main cases.

(1) \emph{Linear impulsive waves with velocity $c$.} For these waves we find general solutions depending on an initial travelling pulse of the e.m. scalar $S$ having arbitrary form. Magnitude order estimates in the case of $\delta$-like initial pulses indicate the generation of gravitational forces per unit mass of the order of $f_x \sim 10^{-29} \psi_0 S_0^2 c^3 \tau$ in SI units, where $\psi_0$ is the unknown VEV of the gravitational scalar $\psi$, and $S_0$ is the initial amplitude of $S$. Possible estimates of $S_0$ have been discussed in Sect.\ \ref{MO}.

(2) \emph{Non-linear soliton-like waves with velocity $v \neq c$.} Their existence requires a very special configuration of the material sources, capable of generating at $x=0$ a precise configuration of the $S$-field compatible with (\ref{Simpli}), and, besides, these sources must be active in the distant past, formally since $t\rightarrow-\infty$. 
They appear to be similar to the general linear impulsive waves, the main difference being that their speed is not $c$. In this respect, it is important to recall that the speed is measured in the Minkowski flat space-time background, and not measured locally. This is reminiscent of a type of Alcubierre space-time solution \cite{alcubierre1994}. In fact it can be checked that depending on the sign of the e.m. scalar coupling $K_\phi$, the soliton-like waves can violate the weak, dominant and strong energy conditions in the case $v>c$ or $v<c$. 

The second kind of experiments concerns small anomalous stationary forces reportedly acting on asymmetric high-frequency resonant e.m. cavities. The most recent data set an experimental upper limit of the order of 10 nN \cite{neunzig2022}.  Our theoretical estimates, Eqs. (\ref{Fzestimated}) and (\ref{fzcav}), indicate that extremely large values of the cavity quality factor and of the electromagnetic power are required to achieve non-negligible effects, in agreement with the absence of measurable forces in the experiments reported. 

More general mathematical forms of these solutions can be found, but this is beyond the scope of this work, whose main purpose was to show the existence of possible new forms of e.m.-gravitational coupling in the framework of extended theories.

\bibliographystyle{ieeetr}
\bibliography{scalar_tensor}

\end{document}